# A Conceptual Framework for the Promotion of Trusted Online Retailing Environment in Saudi Arabia

Rayed AlGhamdi

Faculty of Computing & Information Technology, King Abdulaziz University

PO Box 80200 Jeddah 21589, Kingdom of Saudi Arabia

E-mail: raalghamdi8@kau.edu.sa

Steve Drew

Information & Communication Technology School, Griffith University

Parklands Drive, Southport, QLD 4215, Australia

E-mail: s.drew@griffith.edu.au

Thamer Alhussain

Computer Sciences & Information Technology College, King Faisal University

PO Box 380 Ahsaa 31982, Kingdom of Saudi Arabia

E-mail: thamer_om@hotmail.com

**Abstract**

This paper presents a model conceptual framework that is aimed at promoting trust in the online retailing environment in the Kingdom of Saudi Arabia (KSA). Despite rapid Internet growth, the development of online retailing in Saudi Arabia continues to progress very slowly compared to that of the developed and leading developing countries. To determine the reason behind the sluggish growth of online retailing in the KSA, a mixed methods study involving retailers and customers was conducted in four stages. The outcomes of the study point to distrust in the online retailing environment in Saudi Arabia as a key inhibitory factor for growth. As such, a five-part model is proposed to promote trust in the online shopping environment in the KSA.

## 1. Introduction

The electronic commerce (e-commerce) revolution started in the 90s in much of the developed world. In the near future, this trend will, without doubt, become not only a tool to increase income but an essential means for business competition as well. Since 2000, the rapid growth of e-commerce activities among developed countries has been obvious. Global e-commerce spending is worth about US$10 trillion at present compared to US$0.27 trillion in 2000. The United States, followed by Europe, constitutes the largest share with about 79% of the global e-commerce revenue. The African and Middle East regions, on the other hand, have the smallest share with about 3% of the global e-commerce revenue (Kamaruzaman, Handrich & Sullivan, 2010). While developed nations have become familiar with e-commerce, it is still considered an innovation in Saudi Arabia. Rogers (2003) defined an innovation as an idea, practice or object that is perceived as new by an individual or other unit of adoption.

In Saudi Arabia, e-commerce is still a new wave in the country's information technology revolution. Although Saudi Arabia has the largest and fastest growing Information and Communication Technology (ICT) marketplaces in the Middle East (Saudi Ministry of Commerce, 2001; U.S. Commercial Services, 2008; Alfuraih, 2008), it is still experiencing a very slow growth in e-commerce (Aladwani 2003; CITC 2007; Agamdi 2008). Only a tiny number of Saudi commercial organisations, mostly medium and large companies from the manufacturing sector, are involved in e-commerce activities (CITC, 2010). Sadly, the Kingdom of Saudi Arabia (KSA) is a 'rich' developing nation that does not value Internet commerce highly. Political power and monetary power both reside with the government (and the royal family) and it seems that e-commerce development has been hindered to ensure that no non-government business will compete for monetary power. The KSA's amassed wealth means that competition and efficiencies are not as important as they are in the broadly market driven Western economies. The KSA is still pretty much a 'closed shop' with respect to international commercial trade; therefore, retailers are not enticed to expand their operations to the online marketplace. Arguably, public education regarding the opportunities offered by the Internet is very much needed in order to equip the retailers and shoppers in Saudi Arabia with the knowledge and skills that will make them take advantage of the medium.

With existing literature as a basis, along with our own studies into the particular factors that inhibit stakeholders from embracing e-commerce development in the KSA, we have developed a multi-part framework that will help promote greater trust in the online retail medium. In the next section, we present a brief overview of our study's findings with regards to the KSA's slow adoption of e-commerce. Similarly, we also discuss briefly the method that was used in our research. In addition, we also discuss the analysis of the data, particularly the part where trust is seen as an inhibitory factor for e-commerce growth. Based on the analysis, we present a five-part conceptual framework that will serve as a guide in promoting trust in the online retail environment in Saudi Arabia.

## 2. Review of Previous Studies

Basically, e-commerce is commerce that is enabled through Internet technologies. It includes both pre-sale and post-sale activities (Whiteley, 2000; Chaffey, 2004). E-commerce has various models that can be classified in a variety of ways. The following are five e-commerce models proposed by Laudon and Traver (2007): Business to Customer (B2C), Business to Business (B2B), Customer to Customer (C2C), Peer to Peer (P2P) and Mobile Commerce (M-commerce). Davis and Benamati (2003) categorise e-commerce into B2C, B2B, C2C and Business to Employee (B2E). There are also Business to Government (B2G) and Consumer to Government (C2G) models to consider. The B2G and C2G models have implications to the area of e-government and m-government, which are closely related to e-commerce. This study is focused primarily on online retailing, which is a type of B2C (To & Ngai 2006). Online retailing is an Internet-enabled version of traditional retail. It includes four sub-types: Virtual Merchants (online retail store only), Bricks-and-Clicks e-retailers (online distribution channel for a company that also has a physical store), Catalogue Merchants (online version of direct mail catalogue) and Manufacturers who sell directly over the web (Laudon & Traver 2007).

The authors conducted four sub-studies to determine the factors that might influence retailers to offer online channel sales and for customers to purchase from online retailers in Saudi Arabia. Two studies were conducted with retailers and another two were conducted with customers. A mixed methods research design was used with each research sample. A qualitative approach was conducted first to explore the issues that influence the phenomenon. A quantitative approach was used for testing the qualitative findings in a wider sample.

The qualitative study by AlGhamdi, Drew and AlFaraj (2011) established a list of factors that inhibit customers from purchasing goods online from an e-retailer in the KSA. These inhibitors were found to be: (1) lack of home mailbox; (2) feeling uncomfortable paying online with a credit card; (3) do not know e-retailers in Saudi Arabia; (4) lack of experience in buying online; (5) not having easy and fast access to the Internet; (6) lack of physical inspection of a product; (7) personal information (name, mobile number, e-mail and so on) privacy; (8) lack of clear regulations and legislation for e-commerce in the KSA; (9) lack of language understanding if the website or part of it is in English and (10) not trusting e-retailers in Saudi Arabia. The qualitative study also established a list of incentives that may encourage customers to purchase online from e-retailers in the KSA. These enablers were found to be: (1) competitive prices; (2) owning a home mailbox; (3) easy access and fast speed of the Internet; (4) provision of educational programs; (5) local banks make it easy to own a credit card; (6) professional and easy to understand design of the e-retailer's website, including showing complete specifications with photos of the products; (7) existence of a physical shop apart from the online shop (Brick and click); (8) existence of online payment options other than credit cards and (9) existence of government support, supervision and control. The purpose of this paper is to establish the relative strengths of the inhibiting and enabling factors stated above.

AlGhamdi, Drew and Al-Gaith (2011) also adopted a qualitative approach. They used information obtained from a series of interviews with 16 Saudi retailers to form a list of factors that inhibit or discourage retailers from adopting online retail. These inhibitors were found to be: (1) setup cost; (2) delivery issues; (3) resistance to change; (4) lack of e-commerce experience; (5) poor ICT infrastructure; (6) lack of online payment options to build trust; (7) do not trust online sales; (8) habit/culture of people in Saudi Arabia not favourable toward buying online; (9) lack of clear rules/laws for e-commerce in Saudi Arabia; (10) difficulties in offering a competitive advantage on the Internet; (11) not profitable/not useful and (12) type of business/products are not suitable to be sold online. This qualitative study also established a list of incentives that are likely to enable or encourage retailers to adopt the online channel. These enablers were found to be: (1) develop strong ICT infrastructure; (2) government support and assistance for e-commerce; (3) educational programs and building the awareness of e-commerce; (4) trustworthy and secure online payment options and (5) provision of sample e-commerce software for trialling.

To obtain quantitative indications of the relative strengths of these inhibiting and enabling factors for both customers and retailers, two quantitative studies were conducted (AlGhamdi, Nguyen, Nguyen & Drew, 2011a; AlGhamdi, Nguyen, Nguyen & Drew 2011b). Table 1 in the Results and Discussion section summarizes the findings of these four-stage studies. It also provides an overview of the inhibitors and enablers for customers and retailers ranked according to their strength.

## 3. Research Methodology

The entire study on the online retail in Saudi Arabia was done through the combination of qualitative and quantitative approaches. The mixed methods approach was used on the retailers and customers samples. The qualitative study was conducted first for exploration purposes. This was followed by the quantitative approach that was based on the qualitative findings for testing purpose. This type of approach is called exploratory mixed methods design (Creswell 2008, p. 561) and it is done 'to explore a phenomenon and to [collect] quantitative data to explain relationships found in the qualitative results (Creswell 2008, p. 561).' The mixed methods approach helps to provide an in-depth investigation of the research problem (Morse 2003; Johnson & Onwuegbuzie 2004; Greene 2007; Alise & Teddlie 2010; Feilzer 2010).

Interviews were conducted with 16 Saudi participants (eight males and eight females) aged 16 to 45 years. A qualitative content analysis was used to identify the factors that positively and negatively influenced the customers' decision to purchase from online retailers in Saudi Arabia. A questionnaire survey based on the findings of the qualitative study was used to obtain more information about the relative strengths of these factors. Typically, a question that asked for information about the participant's background and attributes would provide a set of choices, including an open answer (e.g., 'other') where the participant could insert additional information if he or she wished. The two key questions were 'What factors inhibit or discourage you from buying online from e-retailers in Saudi Arabia?' and 'What would enable you to buy online from e-retailers in Saudi Arabia?' The participants were given a list of 11 options to select from for the first question and 10 options for the second (in each case, the last option is 'other reasons'). The respondents were allowed to select as many of the available options as they wish, including the open answer. The survey questions were in Arabic. Two forms were distributed and a total of 680 responses were obtained. Out of the 680, 306 were obtained from the paper surveys and 374 were obtained from the online survey. The sample covered different age groups from 15 to 60+ and it included 320 females and 360 males, all of which came from different cities in Saudi Arabia.

Interviews were also conducted with 16 decision makers from the retailers' group (including owners, headquarter managers, marketing managers and IT managers), which covered various business categories in Saudi Arabia. A qualitative content analysis was used to identify the factors that positively and negatively influenced the retailers to adopt and use the online channel for their business. A questionnaire survey based on the findings of the qualitative study was used to gain more information about the relative strengths of these factors. Typically, a question that asked for information about the participating retailer's attributes would provide a set of choices, including an open answer (e.g., 'other'), where the participant could insert additional information if he or she wished. The two key questions were 'What factors inhibit or discourage your company from implementing an online system to sell on the Internet?' and 'What factors help or encourage your company to implement an online system to sell on the Internet?' The participants were given a list of 13 options to select from for the first question and six options for the second (in each case, the last option is 'other'). The respondents were allowed to select as many of the available options as they wished, including the open answer. The survey questions were designed in English with an Arabic translation version available so that the participant could opt for the most familiar language. Around 200 paper copies of the questionnaire forms were distributed in person to retail businesses in Jeddah (the main economic city in Saudi Arabia), Riyadh (the capital city) and Al-Baha during February and March of 2011. Potential participants were selected via the 'snowballing' approach where some of them were initially approached and then asked to recommend others who might be willing to participate and so on. Judgment was also exercised to ensure that the questionnaire forms were delivered to a wide range of businesses in terms of their size and the type of products or services they offered. A total of 80 completed forms were returned, giving a response rate of around 40%. The authors collected the email addresses of 416 retail companies that were members of the Jeddah and Riyadh chambers of commerce. Associates of the authors provided an additional list of around 50 business email addresses with which they were familiar. Invitations to participate online were sent via email to these 466 addresses, but around 100 were returned because the addresses were invalid. The authors also received Saudi Post's assistance in emailing the 50 retailers that had registered in their e-mall (this e-mall had started in October 2010). Thus, in total, there were 416 retailers who could potentially participate in the online survey via the Griffith website. At the time of writing (May 2011), 68 retailers had participated, implying a response rate of 16.3%.

## 4. Results and Discussion

Table 1 summarizes the findings of our study. It provides an overview of the inhibitors and enablers for the customers and retailers ranked according to their strength. Obviously, the most important factor that affects customers' decisions to purchase from online retailers in the KSA is the lack of trust. The lack of trust in this regard relates to the online environment in the KSA itself. Therefore, lack of trust can be defined here as fair to use the online environment as a means to purchase online because online retailing is new in the KSA and there is a lack of regulation for this new type of business. Furthermore, the KSA's current infrastructure does not support nor encourage retailers and customers to go ahead with this type of business.

As illustrated in Table 1 below, a large number of the customer sample fear that they might not receive their purchased products in the form or quality specified on the website. As such, they are not comfortable to buy without physical inspection. The enabler ranked number one for them is to have a traditional shop alongside an online store so they can visit in case there is a problem. In the qualitative study, this was well explained. The chance of current leading retailers being successful when opening an online channel is high. The customers trust the retailers of leading brand name products and they may have no problem purchasing from these retailers online because they already trust them. In the case of customers having a problem or wanting to return or replace goods, they can visit the retailer's traditional shop to ask for a replacement.

On the other side, current retailers see that the most important factor inhibiting their selling goods online is the existing retail culture of people in Saudi Arabia. With indications that e-retail uptake by customers may be slow, the number of sellers who will go ahead with this new type of business is relatively not high. Therefore, the results suggest a number of factors that may encourage online retail adoption, such as removing their security and trust concerns. For this, offering trustworthy secure online payment options ranked as the number one enabler. Globally, the most popular payment method for online businesses is via credit card. In Saudi Arabia, there is a negative perception involving the inherent risks of using credit cards to pay online. However, this negative perception is changing slowly. Another problem is that the rate of credit card ownership in Saudi Arabia is low compared to developed nations. The reasons for the low adoption of credit cards as mode of payment relate to several issues, such as the desire to avoid paying interest (called 'Rba'), which is forbidden in Islamic law. In addition, some banks require that a minimum amount of money be deposited into the account every month. Further discussion for this point will be provided later.

The above trust concerns are reinforced with the absence of a clear e-commerce law in Saudi Arabia. Interestingly, the inhibitors and enablers that ranked number 2 in both sides (customers and retailers) all relate to e-commerce law and government support. The Saudi government's role in regulating, supporting and facilitating e-commerce activities within the country seems to be missing. While the government has played a major role in promoting the rapid growth of ICT in general, it also appears to have placed rather less emphasis on e-commerce than on e-learning and e-government (AlGhamdi, Drew & Alkhalaf 2011; Alhussain & Drew, 2010; Alfarraj, Nielsen & Vlacic, 2010). The Saudi Ministry of Commerce currently plays no active role in the e-commerce activities within the country (AlGhamdi, Drew & Alkhalaf, 2011). Similarly, the Saudi Ministry of Communications and Information Technology, whose role is to promote e-commerce, is not recognized (CITC, 2010).

We have also found that 62% of the customers in the research sample who have purchased online (AlGhamdi, Nguyen, Nguyen & Drew, 2011a) purchased from international retailers and have no experience purchasing online from Saudi retailers. The reason behind this trend, apart from the competitive prices of course, is because they trust these world renowned retailers. Therefore, what needs to be done in order to promote e-commerce and, in particular, online retailing growth in Saudi Arabia? It is clear here that there is lack of trust in the online shopping environment in Saudi Arabia. Promoting trust in the Saudi online shopping environment should be given high priority. A trusted environment is not only good for customers; it is beneficial for retailers as well as it helps them attract more customers to their business. In this regard, trust is not an issue with the sellers. The issue is with the trust in the online environment. For example, poor delivery system/infrastructure, ICT infrastructure and the non-existence of an e-commerce law all contribute to the distrust in the online environment in terms of conducting purchase transactions. In this paper, developing trust in the online environment needs to be taken care of by policymakers, government and the industry.

In order to promote trust in the Saudi online shopping environment and based on the finding of previous studies, we propose a five-part conceptual model as illustrated in Fig. 1. These five parts are not the only requirements needed; there are also other things that need to be investigated. However, based on our findings, these five parts have high priority and they need to be ensured/applied first before any other recommendations.

**5. The Five-part Conceptual Model**

The five-part model reflects the five factors that contribute most to the promotion of trust in the online shopping environment in Saudi Arabia. The five factors include:

1) Offering Trustworthy and Secure Online Payment Options

2) Consumer Protections

3) Clarifying Marketplace Rules

4) Certification Authority (CA)

5) Strengthen Delivery Systems

Here, we explain in further detail the radial 'wings' that make up the factors that contribute to this model.

*5.1 Offering Trustworthy and Secure Online Payment Options*

In the West, using credit cards for payment is the most popular method in conducting online purchases. In Saudi Arabia, the use of credit card is low compared to developed nations. Many consumers are reluctant to use credit cards online because of lack of trust and because some consumers are culturally averse to carrying out transactions linked with conventional interest rates. Low credit card usage contributes to the low diffusion rate of e-commerce in the country. As such, providing alternative, trustworthy and easy-to-use payment options is a critical need for the industry. Some local banks have started making normal ATM cards that are to be used as debit cards. These cards, however, are limited in their uses and not many banks provide them. Debit cards would present a solution here by allowing customers to control the amount of money placed in their accounts. Other possible solution includes the use of trusted third party payment systems, such as Paypal, which act as link between the credit card holders and the sellers. Systems similar to PayPal help to resolve problems between customers and sellers. PayPal, in this regard, acts as a third party that communicates with both parties to help resolve problems and to reserve rights. Another option is the electronic bill presentment and payment system called SADAD that Saudi Arabia developed for billers and payers who are residents of the country. In essence, SADAD facilitates data exchange between registered billers and the nation's commercial banks. It relies on existing banking channels (such as Internet banking, telephone banking, ATM transactions and even counter transactions) to allow bill payers to view and pay their bills via their banks (for more details, see SADAD 2004). Many consumers are comfortable with using SADAD. However, follow-up comments from some of the respondents in our study suggest that small-to-medium businesses see the initial costs of registration with SADAD and the ongoing transaction processing fees as being too high. Currently, SADAD is limited to 100 billers only. This means that only the biggest billers in Saudi Arabia have access to this system. 'This limits most Internet merchants to using credit cards, short code SMS and prepaid cards as electronic payment methods. To address this limitation, SADAD is currently working on Biller Base Expansion, which will increase its biller options from around 100 to 20,000' (CITC 2010). Thus, cooperation between the government and the industry is needed to provide various credible payment channels for online transactions in Saudi Arabia.

*5.2 Consumer Protection*

Currently, there is a consumer protection body that regulates traditional commerce. However, Saudis are frustrated that this body is not acting as expected and does not cover online commerce. Not only is an online consumer protection body non-existent; it is also compounded by the fact that no national e-commerce law exist either. In this context, consumers are rightly reluctant to purchase online. Consumer protection in electronic commerce is very important and this can be a central part of e-commerce law. It is not enough to include a law that protects consumers' rights (including security of transactions and privacy) on paper. Consumers must also feel confident that there is a body that will represent them along with clear procedures on how to register complaints and seek redress. It is most important for a potential customer to know that there are well defined policies and processes in place, with enforceable expectations and rights, rather than just to know that a consumer protection law exists. For example, a customer purchased a mobile phone from an online retailer in Saudi Arabia and received a mobile phone quite different in specification to the one that was ordered on the website. The retailer refused to accept the mobile phone when the customer wanted to return it. The customer, in this case, would want to know where to file the complaint, how to contact about the complaint and how long it would take to resolve the problem. Clear procedures and fast processing of customers' complaints are factors that ensure customer satisfaction, making them feel that their rights are protected in this type of business.

*5.3 Clarifying Marketplace Rules*

Clarifying online retail marketplace rules is essential to help retailers (and customers) to understand their rights and responsibilities. In Canada, for example, clarifying these rules has been placed as a priority for action in the Canadian e-commerce plan to make Canada a world leader in the development and use of e-commerce by the year 2000 (CEC 1998). These rules include legal and commercial frameworks, financial issues and taxation and the protection of intellectual property.

The law for traditional business in Saudi Arabia is clear. The law for e-commerce in the country, however, is far from clear. Although Saudi Arabia contributes to the efforts of the United Nations Commission into International Trade Laws (UNCITRAL) (Saudi Ministry of Commerce 2001), there is still a need to have major developments in terms of e-commerce regulations. Arguably, legislations and rules to protect the rights of all parties involved in e-commerce transactions are badly needed (Al-Solbi and Mayhew 2005; Agamdi 2008). It is not clear yet whether the Saudi judicial system will accept the adjudication of e-commerce disputes. Most of the transactions done by online businesses are paperless and signatures are digital. Whether or not these current forms of identification will be recognized by the Saudi judicial system is unclear and moves towards the use of public key cryptography to provide binding digital signatures is not yet tried. All of these details need to be clarified before development can progress. However, as explained earlier, no government body in Saudi Arabia actively supports the advent of e-commerce activities in the country. It is apparent that the Ministry of Commerce, which is the most relevant government body, has suffered a diminished role as indicated by the collected data.

Financial and taxation issues may not present a problem in Saudi Arabia because traditional business is a tax-free environment. The same may be applicable for online businesses. The rules of the marketplace may also include the requirements for certifying an organisation's trust status through a system of issuing certification authorities (CAs).

*5.4 Certification Authority (CA)*

A plan for establishing creditability in the online environment is essential since visitors of the website will not buy online unless they trust the company behind the site. Brand name products and leading retailers may not encounter difficulty in gaining the trust of customers. Their reputations help them to build online businesses more quickly and easily than new or unwell-known retailers. New and unwell-known retailers often encounter major difficulties in terms of attracting online customers and getting customers to trust them. These factors negatively affect their decision to open an online channel. To solve this problem, Certification Authorities (CAs) exist. CAs help new and relatively unknown retailers to build trust with new customers. A Certification Authority is 'an entity that issues digital certificates. The digital certificate certifies the ownership of a public key by the named subject of the certificate. This allows others (relying parties) to rely upon signatures or assertions made by the private key that corresponds to the public key that is certified' (Wikipedia).

In 2001, a National Public Key Infrastructure named the National Center for Digital Certification (NCDC) was created in Saudi Arabia. NCDC 'utilizes digital certificates issued by Certification Authorities (CAs) that meet the rules established by the NCDC's governing body, the National Policy Authority (NPA). NCDC Digital Certificates support authentication, digital signature, encryption and non-repudiation services for access and processing of electronic information, documents and transactions' (NCDC 2011).

The role of the NCDC is important in the online environment. However, there is a need to have something more basic, understandable and clear for customers. In the research sample, some retailers suggest that the Saudi Ministry of Commerce take an active role in this regard. According to a retailer, since the Ministry of Commerce is a government entity, there is no doubt that it is trusted more by citizens. Here are some comments that some retailers raised in this regard: 'People in Saudi Arabia have great confidence in anything that comes through the government.' 'They will have more trust if this subject is sponsored by the government.' 'Building trust with customers is a very important issue and requires a good amount of work and activity to be there. There are people who have two to three years experience in the e-commerce business and who are selling very well but are still very far compared to similar but very small businesses in Europe or Asia. We are working now to have certificates from trusted organisations to build trust with our customers. Like this, there should be a certification body from the government itself to say that this company is a certified company by the local government and that you can buy from them. This is a good way to build trust among customers since the government trusts the certified companies.'

The government may issue these certificates to retailers when specific requirements (including all the relevant policies) are met. These requirements should be clearly explained in the marketplace rules.

*5.5 Strengthen Delivery Systems*

If goods are to be reliably delivered to home residences and businesses, there needs to be a recognised system of property addresses. Until 2005, homes in Saudi Arabia have no mail addresses. Compared to the developed world and some developing countries, Saudi Arabia is late in implementing a postal addressing system. In 2005, the new project for addressing and delivering mails to homes and buildings was announced and approved by the Saudi Post. A mail service called 'Wasel' enables the residents of Saudi Arabia to receive all their mail at their residence. The mails are delivered to their mail box free of charge. To avail of this service, the residents of Saudi Arabia have to call the Saudi Post, visit a Saudi post office or complete a form online to get a mailbox with a physical home address. However, this service does not cover all cities of Saudi Arabia. Up until now, it has only been rolled out in the main cities. The number of subscribers of the 'Wasel' service has reached more than half a million so far. This number represents almost 2% of the population who own an individual house mailbox. As this service is still new in Saudi Arabia, it may take time before its use is recognised and adopted by most of the population in the country. While the service of providing individual house mailboxes is in place, the low percentage of population applying to own a mailbox suggests that some efforts are required to investigate the factors that inhibit citizens from using the service. Reasons for the low use of the service might include the lack of awareness of the growing importance of having a mailbox. They might also include the citizens' distrust in the security of receiving their mail through this new service or its providers. A more basic issue is that among the participants in this research, there are those who are not aware of the existence of a service that delivers mails to individual houses or who do not know that it is possible to get direct addresses for their houses with numbers and streets names. In all cases, while the service is there, more efforts are needed to motivate the citizens to own house mailboxes and to find solutions to the problems of mail delivery.

The Saudi post is owned by the government. Private local and global shipment companies (global companies such as DHL, Aramix, Fedex and local) run their businesses in Saudi Arabia but do not deliver to homes due to the lack of an addressing system. In the current context, a customer must leave his/her phone number with the shipment company. When his/her order has arrived, the shipment company then contacts the customer to come and pick up his/her order from the local office. Local companies may deliver to homes for an extra fee. In such a case, the customer has to explain to the driver where his/her home is located.

As discussed above, delivery systems in Saudi Arabia do not support the development of e-commerce in the country. More efforts are needed to strengthen the delivery systems.

## 6. Conclusion

This paper presented a model conceptual framework that is aimed at promoting trust in the online retailing environment in Saudi Arabia. Saudi Arabia is a late adopter of e-commerce and the development of e-commerce in the country is progressing slowly compared to that of the developed and leading developing countries. A multiple stages study has been conducted to figure out the issues that are causing the slow progress of e-commerce in Saudi Arabia. The outcomes of the study point to distrust in the online retailing environment as a key inhibitory factor for growth. Trust, in this regard, is not an issue with the sellers. Rather, the issue is with the online environment of the KSA. For example, poor delivery system/infrastructure, ICT infrastructure and the non-existence of an e-commerce law all contribute to the distrust in the online environment in terms of conducting a purchase. For this reason, a five-part model was proposed to promote trust in the online shopping environment in the KSA. The model included five parts that are needed for the promotion of trust. These five parts are the following: offering trustworthy and secure online payment options, ensuring consumer protection, clarifying marketplace rules, issuing certification authority and strengthen delivery systems.

Table 1. Overview of the Inhibitors and Enablers for the Customers and Retailers

| Rank | Inhibitors to Customers | Inhibitors to Retailers | Enablers for Customers | Enablers for Retailers |
|---|---|---|---|---|
| 1 | 53.8% Lack of physical inspection | (42.6%) Habit/Culture of people to buy is not encouraging | 62.4% Physical shop as well as online shop | (58.1%) Trustworthy and secure online payment options |
| 2 | 52.4% Lack of clear regulations and legislation | (35.8%) Lack of clear legislations and rules of e-commerce in the KSA | 52.6% Government supervision and control | (53.1%) Government support and assistance for e-commerce |
| 3 | 44.7% Lack of online purchase experience | (35.8%) Lack of e-commerce experience | 50.0% Competitive prices | (39.9%) Develop strong ICT infrastructure |
| 4 | 41.0% Don't trust that personal info will remain private | (25.0%) Products are not suitable to be sold online | 41.5% Trustworthy and secure online payment options | (31.1%) Educational programs and building awareness on e-commerce |
| 5 | 36.6% Do not know e-retailers in Saudi Arabia | (22.4%) Poor ICT infrastructure | 36.2% Owning a house mailbox | (25.7%) Provision of sample e-commerce software to trial |
| 6 | 33.4% E-retailers are not trusted in terms of mailing products in quality same as specified | (20.9%) Lack of online payment options help to build trust | 33.8% Well-designed retailer websites (photos of products) | (15.5%) Other |
| 7 | 31.8% Lack of mailbox for home | (16.9%) Resistance to change | 32.6% Easy access and fast Internet | |
| 8 | 26.3% Uncomfortable paying online using credit cards | (13.6%) Other | 29.0% Provision of educational programs | |
| 9 | 21.9% Lack of English language understanding | (12.8%) We do not trust online sales | 20.6% Local banks make owning credit cards easier | |
| 10 | 7.5% Difficult access to the Internet and slow speed | (10.8%) Delivery issues | 7.5% Other | |
| 11 | 6.0% Other | (10.1%) Setup cost | | |
| 12 | | (7.4%) Cannot offer competitive advantage | | |
| 13 | | (8.1%) Not profitable | | |

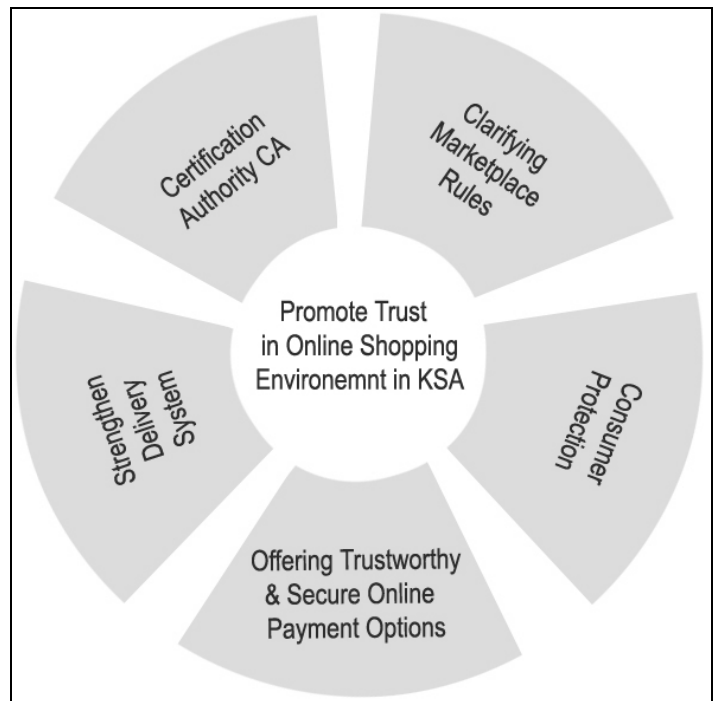

Figure 1. The five-part model that promotes trust in the online retailing environment in the KSA